\documentclass[aps,prb,reprint,superscriptaddress,amsmath,amssymb,showpacs]{revtex4-1}%

\newcommand{\braket}[2]{\left< #1|#2 \right>}

\makeatletter
\newcommand*{\rom}[1]{\expandafter\@slowromancap \romannumeral #1@}
\makeatother

\renewcommand{\AA}{\mathcal{A}}
\newcommand{\CC}{\mathcal{C}}
\newcommand{\BB}{\mathcal{B}}
\newcommand{\DD}{\mathcal{D}}
\newcommand{\LL}{\mathcal{L}}
\newcommand{\dd}{\mathrm{d}}

\newcommand{\ii}{\text{i}}
\newcommand{\Tr}[1]{\text{Tr} \left\{ #1 \right\}}
\newcommand{\tr}[2]{\text{Tr}_{ #1 } \left\{ #2 \right\}}
\newcommand{\av}[1]{\left\langle #1 \right\rangle}

\newcommand{\expo}[1]{\text{exp}\left( #1 \right)}
\newcommand{\e}{\text{e}}
\newcommand{\tS}{\text{S}}

\newcommand{\pdagger}{{\phantom{\dagger}}}
\renewcommand{\bm}[1]{\textbf{\textit{#1}}} 

\newcommand{\power}{P}
\newcommand{\vibpop}{\av{a^\dagger a}}
\newcommand{\pos}{\av{a^\dagger {+}a}}
\newcommand{\elpop}{\av{d^\dagger d}}
\renewcommand{\t}{\text}

\usepackage{braket}
\usepackage{import}
\usepackage{subfigure}
\usepackage{graphicx}
\usepackage{xcolor}
\usepackage{dsfont}
\usepackage{pstricks}
\usepackage{mathrsfs}
\usepackage{bbm}
\usepackage{nicefrac}

\definecolor{plotgreen}{RGB}{0,158,115}
\definecolor{plotblue}{RGB}{0,114,178}
\definecolor{plotorange}{RGB}{230,159,0}
\definecolor{frBlueCMYK}{cmyk}{1,.75,0,0}
\definecolor{r}{cmyk}{0.2,1.,0,0}        

\begin{document}

\title{
Periodically driven open quantum systems with vibronic interaction: Resonance effects and vibrationally mediated decoupling
}

\author{Jakob B\"atge}
\email{jakob.baetge@physik.uni-freiburg.de}
\affiliation{Institute of Physics, University of Freiburg, Hermann-Herder-Str. 3, 79104 Freiburg, Germany}

\author{Yu Wang}
\affiliation{Department of Chemistry, School of Science, Westlake University, Hangzhou, Zhejiang 310024, China }
\affiliation{Department of Physics, School of Science, Westlake University, Hangzhou, Zhejiang 310024, China }
\affiliation{Institute of Natural Sciences, Westlake Institute for Advanced Study, Hangzhou, Zhejiang 310024, China}

\author{Amikam Levy} 
\affiliation{Department of Chemistry, Bar-Ilan University, Ramat-Gan 52900, Israel }

\author{Wenjie Dou}
\email{douwenjie@westlake.edu.cn} 
\affiliation{Department of Chemistry, School of Science, Westlake University, Hangzhou, Zhejiang 310024, China }
\affiliation{Department of Physics, School of Science, Westlake University, Hangzhou, Zhejiang 310024, China }
\affiliation{Institute of Natural Sciences, Westlake Institute for Advanced Study, Hangzhou, Zhejiang 310024, China}

\author{Michael Thoss}
\email{michael.thoss@physik.uni-freiburg.de}
\affiliation{Institute of Physics, University of Freiburg, Hermann-Herder-Str. 3, 79104 Freiburg, Germany}

\begin{abstract}
Periodic driving and Floquet engineering have emerged as invaluable tools for controlling and uncovering novel phenomena in quantum systems. In this study, we adopt these methods to manipulate nonequilibrium processes within electronic-vibronic open quantum systems. Through resonance mechanisms and by focusing on the limit-cycle dynamics and quantum thermodynamic properties, we illustrate the intricate interplay between the driving field and vibronic states and its overall influence on the electronic system. Specifically, we observe an effective decoupling of the electronic system from the periodic driving at specific frequencies, a phenomenon that is mediated by the vibrational mode interaction. Additionally, we engineer the driving field to obtain a partial removal of the Franck-Condon blockade. These insights hold promise for efficient charge current control. Our results are obtained from numerically exact calculations of the hierarchical equations of motion and further analyzed by a time-periodic master equation approach.

\end{abstract}
 
\maketitle

\section{Introduction}

The idea to realize functional devices based on open quantum systems with time-dependently controlled parameters was established by Alicki and Kosloff~\cite{alicki1979quantum,Kosloff1984}. Among various applications, high-efficiency energy converters are particularly targeted~\cite{kosloff2014quantum,Myers2022, cangemi2023quantum}.
Even though the theoretical description is challenging, various model systems have been considered in this context~\cite{Geva1992,Grifoni1998,kosloff2014quantum,Brandner2016,Karimi2016,Newman2017,Kloc2019,Park2019,Koyanagi2022,Anto2023,cangemi2020violation}.
Typical investigations considered specific models that obey an exact solution~\cite{Proesmans2016,Ludovico2016,Restrepo2019,Wiedmann2020} or applied additional approximations.
Examples of such approximations are a weak system-environment coupling approximation~\cite{Kohler2005,Zhou2015,Liu2021,Lu2022}, the limit of a slow variation of the parameters with time~\cite{Kim2002,Moskalets2002,Moskalets2004,Esposito2010,Avron2012,Cavina2017}, the opposite limit of a fast change of the parameters with time~\cite{Moskalets2002,Cavina2021}, or the regime of linear response~\cite{Brandner2016,Brandner2017}.
Moreover, the time-periodicity was utilized via the Floquet formalism to focus on the operation of the system at the limit of long times~\cite{Grifoni1998,Kim2002,Moskalets2002,Kohler2005,Agarwal2007,Ludovico2016II}. 
Electronic-vibrationally interacting systems, in particular, have been studied in the context of autonomous driving without external control\cite{Novotny2003,Strasberg2021,Waechtler2022}, under the influence of an AC bias voltage\cite{Kuperman2020,Preston2020,Kuperman2022}, with an optical driving field~\cite{Lehmann2004,Franco2008,May2008II}, and with a time-dependent electronic state energy~\cite{Zhou2015}.

In this paper, we employ the numerically exact hierarchical equations of motion (HEOM) approach\cite{Tanimura1989,Tanimura1990,Jin2008,Baetge2021} to describe the dynamics of periodically driven open quantum systems with electronic-vibrational interaction.
Thereby, we employ our previously introduced framework~\cite{Baetge2022} for the intriguing case of periodic driving. 
Such periodic protocols are of particular importance for the replication of conventional thermodynamic cycles in quantum systems. We, thus, turn towards electronic-vibrationally interacting systems as functional devices under out-of-equilibrium conditions.
We emphasize that the HEOM approach is applicable for a wide range of driving frequencies and can resolve both the short- as well as the long-time dynamics induced by the driving.
Here, we limit the periodic driving to the energy of the electronic state, which can be realized by an externally controlled gate voltage. However, the approach can be straightforwardly extended to more time-dependent parameters without physical limitations.
\begin{figure}[!b]
    \centering
    \includegraphics[width=7.9cm]{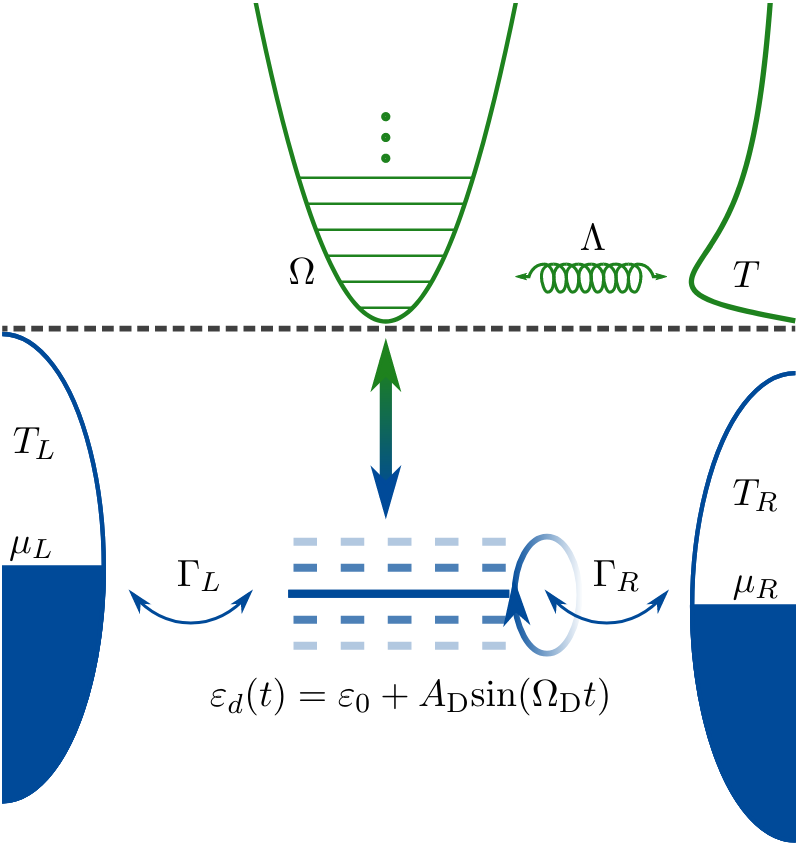}
    \caption{Sketch of the model consisting of a periodically driven electronic state interacting with a single vibrational mode and two electronic reservoirs. The vibrational mode is further coupled to a vibrational heat bath.}
    \label{fig:periodic_system_sketch}
\end{figure}
We find that observables such as the induced power, electronic population, and charge current show strong resonance effects when the driving frequency equals an integer multiple of the vibrational frequency.
In particular, we observe a resonance mechanism in which the vibrational mode counteracts the effect of external driving on the electronic system. 
Moreover, external driving can trigger a removal of the Franck-Condon blockade. These insights hold promise for efficient charge current control.
A comparison of the results of the HEOM method and the Floquet quantum master equation yields qualitative agreement. 
Overall, we demonstrate the power of using HEOM to study dynamics and thermodynamics for open quantum systems subjected to time-dependent driving.

The paper is organized as follows:
In Sec.~\ref{sec:ModelandMethodology}, we present the details of the periodically driven model, and we briefly introduce both the HEOM approach and the Floquet quantum master equation induced by a periodic driving field. We also introduce definitions for the description of the driving-induced dynamics.
Then, in Sec.~\ref{sec:interplay_driving_field_harmonic_mode}, the periodically driven electronic-vibrationally interacting system is examined, where the effects of bias voltages and vibrational relaxation are added layer by layer.
Finally, we conclude in Sec.~\ref{sec:Conclusion}.

\section{Model and Methods}
\label{sec:ModelandMethodology}

\subsection{Model}
We consider the archetypal electronic-vibrational transport model with the inclusion of an external periodic driving field, as sketched in Fig.~\ref{fig:periodic_system_sketch}. 
Using units with $\hbar=k_\text{B}=1$, the Hamiltonian reads
\label{eq:Hamiltonian}
\begin{align}
 H =& \varepsilon_d(t)d^\dagger d + \lambda (a^\dagger +a) d^\dagger d + \Omega a^\dagger a \label{eq:systemHamiltonian} \nonumber \\
 &+\sum_{k\alpha}\nu_{k\alpha} \left(c^\dagger_{k\alpha}d + d^\dagger_{\phantom{k\alpha}} c_{k\alpha}^{\phantom \dagger}\right) + \sum_{k\alpha}\varepsilon_{k_\alpha} c_{k\alpha}^\dagger c_{k\alpha} \nonumber\\
 &+(a^\dagger +a)\sum_j \xi_j (b_j^\dagger + b_j) + \sum_j \omega_j b_j^\dagger b_j^{\phantom \dagger} \nonumber\\
 &+(a^\dagger +a)^2\sum_j\frac{\xi_j^2}{\omega_j}.
\end{align}

Here, the system part of the Hamiltonian consists of an electronic state with an externally controlled energy, 
$\varepsilon_d(t)$, a harmonic mode with frequency, $\Omega$, and an adiabatic coupling of the electronic state to the harmonic mode with coupling strength, $\lambda$. Both the electronic state and the harmonic mode are addressed by their creation (annihilation) operator $d^\dagger(d)$ and $a^\dagger(a)$.
The externally controlled electronic state energy is further assumed to obey a sinusoidal time-dependence, $\varepsilon_d(t)=\varepsilon_0 + A_\text{D}\text{sin}\left(\Omega_\text{D} t\right)$, where $A_\text{D}$ denotes the amplitude of the driving field and $\Omega_\text{D}$ characterizes the driving frequency. The system part of the Hamiltonian can be diagonalized via the small polaron transformation,~\cite{Mahan2013} where the electronic-vibrational interaction results in the renormalized electronic energy,
\begin{equation}\label{eq41}
\begin{split}
    \bar{\varepsilon}_d (t) \equiv \varepsilon_d(t) - \frac{\lambda^2}{\Omega}.
\end{split}
\end{equation}

Electronic transport through the system is introduced by two macroscopic electron reservoirs acting as electron sources and drains. When isolated, both electron reservoirs, $\alpha \in \{L,R\}$, are characterized by their chemical potential, $\mu_\alpha$, and temperature, $T_\alpha$. Microscopically the electron reservoirs are described by electronic states with an energy $\varepsilon_{k\alpha}$, which are addressed by the corresponding creation (annihilation) operators $c^{\dagger}_{k\alpha}(c^{\phantom \dagger}_{k \alpha})$. The collective influence of each electronic reservoir is further characterized by its spectral density
\begin{align}
\Gamma_\alpha(\varepsilon)=2\pi\sum_k|\nu_{k\alpha}|^2\delta(\varepsilon-\varepsilon_{k\alpha})=\Gamma_\alpha \frac{D_\alpha^2}{D_\alpha^2+(\varepsilon-\mu_\alpha)^2}.
\end{align}
Here, $\Gamma_\alpha$ denotes the coupling strength of the electronic reservoir and $D_\alpha$ the bandwidth of the Lorentzian-shaped spectral density. In the following, we effectively investigate wide bands by the choice of $D_\alpha=30\,$eV and assume symmetrically coupled reservoirs with $\Gamma_L=\Gamma_R=\frac{\Gamma}{2}$.

The environment also introduces vibrational relaxation to the system. The macroscopic heat bath with temperature $T$ is microscopically characterized by a continuum of mutually independent harmonic modes, where the $j$-th mode has a frequency $\omega_j$, the creation (annihilation) operator $b_j^\dagger(b_j^{\phantom\dagger})$, and the coupling strengths $\xi_j$. The collective influence of the microscopic vibrational modes is specified by the Ohmic spectral density
\begin{align}
    \Lambda(\omega)=\pi\sum_j|\xi_j|^2 \delta(\omega-\omega_j)=\Lambda \frac{\omega}{\Omega}\frac{\omega_c^2}{\omega_c^2+\omega^2}.
\end{align}
Here, the $\Lambda$ denotes the collective coupling strength and $\omega_c$ is the cut-off frequency of the Lorentzian cut-off. Throughout this paper, we choose $\omega_c=\Omega$. Furthermore, the last term in Eq.~\eqref{eq:systemHamiltonian} counteracts the renormalization of the harmonic oscillator frequency induced by the coupling to the environment~\cite{Caldeira1983,Grabert1988}.

\subsection{Hierarchical equations of motion (HEOM) method}

In the following, we present the most important steps of the derivation of the numerically exact HEOM approach for the model under investigation.
Thereby, we closely follow Refs.~\onlinecite{Schinabeck2016} and \onlinecite{Baetge2021}. 
More detailed derivations are presented in Refs.~\onlinecite{Tanimura1989,Jin2008}.  

The derivation of the HEOM is based on the system-environment partitioning (see Eq.~\eqref{eq:Hamiltonian}).
The central quantity of the approach is the reduced density matrix $\rho(t)$ of the system, where the bath degrees of freedom are traced out.
The influence of the environment on the system dynamics is taken into account by the Feynman-Vernon influence functional without approximations.
For our model Hamiltonian, all information about system-environment coupling is completely encoded in the two-time correlation functions of the free environments 
 \begin{subequations}  
 \begin{align}
    \tilde{C}(t-\tau)=&\sum\limits_{j} |\xi^\pdagger_{j}|^2 \av{   {b}_{j}^\dagger (t) {b}_{j }^\pdagger(\tau)  +  {b}_{j }^\pdagger(t) {b}_{j}^\dagger(\tau) } ,
    \\
    C^s_{\alpha}(t-\tau)=&\sum\limits_{k } |\nu_{k \alpha}|^2 \av{ 
 c^{s}_{k \alpha}(t)  c^{\bar{s}}_{k \alpha}(\tau) } ,
 \end{align}
 \end{subequations}  
 which are determined by the spectral densities
\begin{align}
  \tilde{C}(t) = & \int_0^\infty d\omega\frac{\Lambda(\omega)}{\pi} \left[\text{coth}\left(\frac{\beta \omega}{2}\right)\text{cos}(\omega t) - i \text{ sin}(\omega t)\right],
      \\
  C^s_{\alpha} (t)=&\frac{1}{2 \pi} \int_{-\infty}^\infty \dd \varepsilon\, \e^{s \ii \varepsilon t} \Gamma_{\alpha} (\varepsilon) f (s (\varepsilon-\mu_\alpha)).
\label{eq:C_FT}
\end{align}
Here, $f(\varepsilon)=\left( \expo{\beta \varepsilon} +1 \right)^{-1}$ denotes the Fermi-Dirac distribution function and $\beta=T^{-1}$ the inverse temperature (with $ k_\text{B}=1$).
Furthermore, the notations $c^+ \equiv c^\dagger$, $c^- \equiv c$ and $\bar{s}\equiv -s$ are employed.
To derive a closed set of equations of motion within the HEOM formalism, all correlation functions are expressed by sums over exponentials~\cite{Jin2008}.
To this end, the Fermi, as well as the Bose distribution,  are represented by sum-over-poles schemes employing Pad\'e decompositions~\cite{Ozaki2007,Hu2010,Hu2011}. Recently, more advanced representations have been presented extending the applicability of the approach~\cite{Chen2022,Xu2022}.
Accordingly, the correlation functions of the free baths are given by 
\begin{subequations}
    \begin{align}
        \tilde{C}(t)=&\Lambda \sum_{p=0}^{p_\text{max}} \tilde{\eta}_p \e^{-\tilde{\gamma}_p t},
        \\
        C^s_{\alpha}(t)=&\Gamma_\alpha \sum_{q=0}^{q_\text{max}} \eta_{\alpha,q}  \e^{-\gamma_{\alpha,s,q} t}.
    \end{align}
\end{subequations} 
Therefore one obtains the HEOM in the form of
\begin{alignat}{2}
   \frac{\partial}{\partial t} \rho^{(m|n)}_{\bm{g}|\bm{h}} =& -\left( i \LL_\tS + \sum_{l=1}^m \tilde{\gamma}_{g_l} + \sum_{l=1}^n \gamma_{h_l} \right) \rho^{(m|n)}_{\bm{g}|\bm{h}}
   \nonumber \\
   &- \sum_{h_x} \AA_{h_x} \rho^{(m|n+1)}_{\bm{g}|\bm{h}^+_x} -\sum_{l=1}^n (-1)^l \CC_{ h_l}  \rho^{(m|n-1)}_{\bm{g}|\bm{h}^-_l}
   \nonumber \\
   &+ \sum_{g_x} \BB_{g_x} \rho^{(m+1|n)}_{\bm{g}^+_x|\bm{h}} +\sum_{l=1}^m \DD_{ g_l}  \rho^{(m-1|n)}_{\bm{g}^-_l|\bm{h}},
   \label{eq:general_HEOM}
\end{alignat}
with the multi-indices $g=(p)$ and $h=(\alpha,s,q)$, the notation for the multi-index vectors ${\bm{v}} = {v_1 {\cdot}{\cdot}{\cdot}v_p}$, ${\bm{v}^+_x}= {v_1 {\cdot}{\cdot}{\cdot}v_p v_x}$, and ${\bm{v}^-_l} = {v_1 {\cdot}{\cdot}{\cdot}v_{l-1}v_{l+1}{\cdot}{\cdot}{\cdot}v_p}$,
and the operator $\LL_\tS O = [H_\tS(t),O]$.
In addition, the superoperators $\AA_{h}$, $\CC_{h}$, $\BB_{g}$ and $\DD_{g}$ read
 \begin{subequations}  
\begin{align}
 \AA_{h} \rho^{(m|n)}_{\bm{g}|\bm{h}}=&  \Gamma_{\alpha_h} \left( d^{s_{h}}  \rho^{(m|n)}_{\bm{g}|\bm{h}} + (-1)^{(n)} \rho^{(m|n)}_{\bm{g}|\bm{h}} d^{s_{h}} \right),\\
 \BB_{g} \rho^{(m|n)}_{\bm{g}|\bm{h}}=&  \Lambda  \left[ \left(a^\dagger +a\right) , \rho^{(m|n)}_{\bm{g}|\bm{h}}\right],\\
 \CC_{h} \rho^{(m|n)}_{\bm{g}|\bm{h}}=&\, (-1)^{n} \eta_{h}   {d^{\bar{s}_{h}} }  \rho^{(m|n)}_{\bm{g}|\bm{h}} - \eta^*_{ \bar{h}}  \rho^{(m|n)}_{\bm{g}|\bm{h}} {d}^{\bar{s}_{h}}, \\
 \DD_{g} \rho^{(m|n)}_{\bm{g}|\bm{h}}=&\, \tilde{\eta}_{g}   \left(a^\dagger +a\right)  \rho^{(m|n)}_{\bm{g}|\bm{h}} - \tilde{\eta}^*_{ g}  \rho^{(m|n)}_{\bm{g}|\bm{h}} \left(a^\dagger +a\right).
 \label{eq:general_HEOM_upbuilding_operators}
\end{align}
 \end{subequations}  
Due to system-environment interaction, these superoperators couple the different levels of the hierarchy.

Here, $\rho^{(0|0)} \equiv \rho$ represents the reduced density matrix and $\rho^{(m|n)}_{\bm{g}|\bm{h}}$ $(n+m>0)$ denote auxiliary density matrices, which describe environment-related observables such as the charge current
\begin{alignat}{2}
 I_{\alpha} &= - e \av{\frac{d N_\alpha }{d t}} &&=  e\,\Gamma_{\alpha} \sum_{h_\alpha} s_h \Tr{ d^{\bar{s}_{h_\alpha}} \rho^{(0|1)}_{\pdagger|{h_\alpha}} }.
\end{alignat}
In the following, we consider the symmetrized charge current through the system defined by $I=\frac{I_L-I_R}{2}$.

The importance of the auxiliary density operators to the system dynamics is estimated by assigning them the importance values,\cite{Haertle2013,Baetge2021} 
 \begin{align}
	 \mathcal{I} \left( \rho^{(m|n)}_{\bm{g}|\bm{h}}\right) =&  \left|\prod\limits_{l=1}^{n }\frac{\Gamma}{\sum\limits_{a\in\{1..{l}\}}\hspace{-.2cm} \text{Re}\left[\gamma_{h_{a}}\right]}   \frac{\eta_{h_{l}}}{\text{Re}\left[\gamma_{h_{l}}\right]} \right| \nonumber 
	 \\ & \times \left| \prod\limits_{l=1}^{m }\frac{\Lambda}{\sum\limits_{a\in\{1..{l}\}}\hspace{-.2cm} \text{Re}\left[\gamma_{g_{a}}\right]}  \frac{\eta_{g_{l}}}{\text{Re}\left[\gamma_{g_{l}}\right]} \right|.
	 \label{eq:importance_estimate}
\end{align}
In the calculations presented in this paper, the results are quantitatively converged for truncation of the hierarchy at level $m=2$ and $n=2$, neglecting auxiliary density operators having an importance value smaller than $10^{-9}$.

\subsection{Time-periodic Born-Markov quantum master equation}
Next, we present the time-periodic Born-Markov quantum master equation (QME) to the system above while excluding the vibrational relaxation, i.e., $\Lambda=0$. 
More details about the derivation of the QME can be found in Ref.~\onlinecite{wang2023nonadiabatic}. In this Floquet QME formalism,  the  reduced density matrix of the system is represented by 
\begin{subequations}    
\begin{align}
    \rho_0 = \tr{\text{el}}{\rho d d^\dagger},\\
    \rho_1 = \tr{\text{el}}{\rho d^\dagger d},
\end{align}
\end{subequations}
where $\rho_0$ and $\rho_1$ denote the reduced density matrix with the system electronic level being unoccupied (state 0) and occupied (state 1), respectively. In the limit of weak system-bath couplings, the reduced density matrices evolve as:
\begin{equation}\label{eq35}
\begin{split} 
    \frac{d\rho_0}{dt} = & -i[h_0,\rho_0]-\sum_k|\nu_k|^2 \int_0^\infty d\tau \times \\ &
     \Big[ e^{i\varepsilon_k\tau-i(g(t)-g(t-\tau))}f(\varepsilon_k) e^{-ih_1\tau}e^{ih_0\tau}\rho_0 \\ & - e^{i\varepsilon_k\tau-i(g(t)-g(t-\tau))}(1-f(\varepsilon_k)) \rho_1 e^{-ih_1\tau}e^{ih_0\tau} \\ &
    + e^{-i\varepsilon_k\tau+i(g(t)-g(t-\tau))}f(\varepsilon_k) \rho_0 e^{-ih_0\tau}e^{ih_1\tau} \\ &
    - e^{-i\varepsilon_k\tau+i(g(t)-g(t-\tau))}(1-f(\varepsilon_k))  e^{-ih_0\tau}e^{ih_1\tau} \rho_1 \Big],
\end{split}
\end{equation}
\begin{equation}\label{eq36}
\begin{split} 
    \frac{d\rho_1}{dt} = & -i[h_1,\rho_1]-\sum_k|\nu_k|^2 \int_0^\infty d\tau \times \\ &
     \Big[ e^{-i\varepsilon_k\tau+i(g(t)-g(t-\tau))}(1-f(\varepsilon_k)) e^{-ih_0\tau}e^{ih_1\tau}\rho_1 \\ &
    - e^{-i\varepsilon_k\tau+i(g(t)-g(t-\tau))}f(\varepsilon_k) \rho_0 e^{-ih_0\tau}e^{ih_1\tau} \\ &
    + e^{i\varepsilon_k\tau-i(g(t)-g(t-\tau))}(1-f(\varepsilon_k)) \rho_1 e^{-ih_1\tau}e^{ih_0\tau} \\ &
    - e^{i\varepsilon_k\tau-i(g(t)-g(t-\tau))} f(\varepsilon_k)  e^{-ih_1\tau}e^{ih_0\tau} \rho_0 \Big].
\end{split}
\end{equation}
Here, we used the time-independent part of the system Hamiltonian $H_S = h_0 d d^\dagger + h_1  d^\dagger d$ and introduced $g(t)=\frac{A_D}{\Omega_D}(1-\text{cos}(\Omega_D)t)$. 
In addition, we can employ the Jacobi-Anger expansion to express the term $e^{-ig(t)}$ in the above equations:
\begin{equation}
\begin{split}
    e^{-ig(t)} \hspace{-0.08cm}= e^{-i \frac{A_D}{\Omega_D}}e^{i\frac{A_D}{\Omega_D}\cos{(\Omega_D t)}}\hspace{-0.08cm} = e^{-i\frac{A_D}{\Omega_D}}\hspace{-0.15cm}\sum_{n=-\infty}^{+\infty}\hspace{-0.2cm}i^n J_n(z)e^{in\Omega_D t},
\end{split}
\end{equation}
where $n$ is an integer, $J_n(z)$ is the $n$-th Bessel function of the first kind, and $z=\frac{A_D}{\Omega_D}$. Thus, we can expand the term $e^{-i(g(t)-g(t-\tau))}$ as 
\begin{equation}\label{eq37}
\begin{split}
    e^{-i(g(t)-g(t-\tau))} &= \sum_{n,m}i^{n-m} J_n(z) J_m(z) e^{i(n-m)\Omega_D t}e^{im\Omega_D\tau}.
\end{split}
\end{equation}
Next, we expand the reduced density matrix in a basis of harmonic oscillator eigenstates, ($h_0|i\rangle = E_0(i) |i\rangle$, $h_1|i'\rangle = E_1(i') |i'\rangle$, $\rho_0(i,j)=\left\langle i \right| \rho_0 \left|j\right\rangle$, $\rho_1(i',j')=\left\langle i' \right| \rho_1 \left|j'\right\rangle$), and obtain
\begin{equation}\label{eq39}
\begin{split}
    \frac{d\rho_0(i,j)}{dt} &= -i(E_0(i)-E_0(j))\rho_0(i,j) \\ & \hspace{-0.85cm}
    - \frac{\Gamma}{2}\sum_{i',k}\tilde{f}(E_1(i')-E_0(k))F_{i\rightarrow i'}F_{k\rightarrow i'}\rho_0(k,j) \\ &\hspace{-0.85cm}
    + \frac{\Gamma}{2}\sum_{i',j'}(1-\tilde{f}^{*}(E_1(j')-E_0(j)))F_{i\rightarrow i'}F_{j\rightarrow j'}\rho_1(i',j') \\ &\hspace{-0.85cm}
    - \frac{\Gamma}{2}\sum_{i',k}\rho_0(i,k)\tilde{f}^{*}(E_1(i')-E_0(k))F_{j\rightarrow i'}F_{k\rightarrow i'} \\ &\hspace{-0.85cm}
    + \frac{\Gamma}{2}\sum_{i',j'}\rho_1(i',j')(1-\tilde{f}(E_1(i')-E_0(i)))F_{i\rightarrow i'}F_{j\rightarrow j'},
\end{split}
\end{equation}
\begin{equation}\label{eq40}
\begin{split}
    \frac{d\rho_1(i',j')}{dt} & = -i(E_1(i')-E_1(j'))\rho_1(i',j') \\ &\hspace{-0.85cm}
    - \frac{\Gamma}{2}\sum_{i,k'}(1-\tilde{f}^{*}(E_1(k')-E_0(i)))F_{i\rightarrow i'}F_{i\rightarrow k'}\rho_1(k',j') \\ &\hspace{-0.85cm}
    + \frac{\Gamma}{2}\sum_{i,j}\tilde{f}(E_1(j')-E_0(j))F_{i\rightarrow i'}F_{j\rightarrow j'}\rho_0(i,j) \\ &\hspace{-0.85cm}
    - \frac{\Gamma}{2}\sum_{i,k'}\rho_1(i',k')(1-\tilde{f}(E_1(k')-E_0(i)))F_{i\rightarrow j'}F_{i\rightarrow k'} \\ &\hspace{-0.85cm}
    + \frac{\Gamma}{2}\sum_{i,j}\rho_0(i,j)\tilde{f}^{*}(E_1(i')-E_0(i))F_{i\rightarrow i'}F_{j\rightarrow j'}.
\end{split}
\end{equation}
The above equations are referred to as our time-periodic Born-Markov QME. Here, $E_0(i)=\Omega(i+\frac{1}{2})$ and $E_1(i')=\Omega(i'+\frac{1}{2})+\bar{\varepsilon}_d$, and 
$F$ is the Franck-Condon factor,
\begin{equation}\label{eq43}
\begin{split}
    F_{i\rightarrow i'} &= \braket{i'|i}\\
    &= \sqrt{\frac{p!}{Q!}}\left(\frac{\lambda}{\Omega}\right)^{\hspace{-0.1cm}Q-p}\hspace{-0.2cm}e^{-\frac{\lambda^2}{2\Omega^2}}L_p^{Q-p}\hspace{-0.1cm}\left(\frac{\lambda^2}{\Omega^2}\right)[{\rm sgn}(i'-i)]^{i-i'},
\end{split}
\end{equation}
with $p(Q)$ denoting the minimum (maximum) of $i$ and $i'$, 
and $L_n^m$ is a generalized Laguerre polynomial. We further defined $\tilde{f}(x)$ as a modified Fermi function with Floquet replicas,  
\begin{equation}\label{eq44}
\begin{split}
    \tilde{f}(x) = \sum_{n,m}i^n(-i)^m J_n(z) J_m(z) e^{i(n-m)\Omega_D t} f(x-m\Omega_D),
\end{split}
\end{equation}
which is a time dependent complex number, and $\tilde{f}^{*}(x)$ its complex conjugate.

\subsection{The limit-cycle and its characterization}

\begin{figure}[!t]
    \centering
    \includegraphics{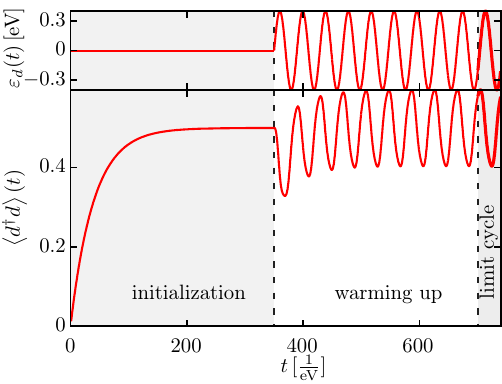}
    \vspace{-0.15cm}
    \caption{Realization of the time-dependent protocol. 
    Shown are the sinusoidal driving and the electronic occupation at different phases of the protocol, until arriving at the limit cycle. The parameters are:
   $T =0.025\,$eV$=\Gamma$, $A_\text{D}=0.4\,$eV, and $\Omega_\text{D}=0.16\,$eV.}
    \label{fig:periodic_procedure}
\end{figure}

The periodic driving field has a significant influence on the dynamics of the open quantum system. In Fig.~\ref{fig:periodic_procedure}, we exemplify our protocol, which is the basis for the subsequent investigations.
The protocol consists of three phases:
\begin{enumerate}
    \item Initialization: We propagate the system according to a constant electronic state energy $\varepsilon_d(t)=\varepsilon_0$ towards its stationary state, which corresponds to the limit $A_\text{D}\to0$.  Thereby we also obtain an estimate for the relaxation time of the system independent of the driving frequency.
    \item Warm-up: We turn on the periodic driving field and record the time-evolution of the system. Once the time-evolution exhibits truly the same periodicity as the driving field, the so-called limit cycle is reached.
    \item Limit cycle: We follow one complete limit cycle with a fine time resolution with a constant increment given by a fraction of the driving period. 
\end{enumerate}
As the limit cycle is of main interest when exploring the operation of periodically driven devices, in this work, we focus on the open-quantum-system response by means of the limit cycle dynamics.

\begin{figure}[!b]
    \centering
    \includegraphics{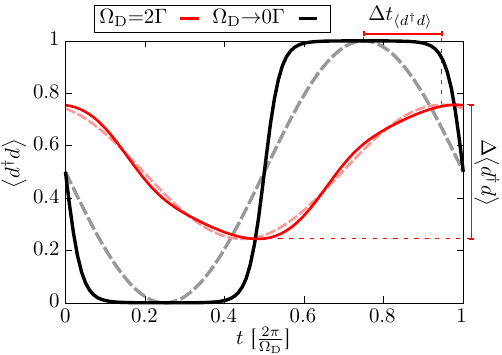}
    \caption{Electronic population dynamics during a limit cycle in comparison to the dynamics in the limit of $\Omega_\text{D}\to0$. On the one hand, the definition of the amplitude $\Delta{\elpop}$ is visualized. On the other hand, the definition of the delay time $\Delta t_{\elpop}$ via the delay time of the Fourier component corresponding to the driving frequency (indicated by the lighter dashed lines) is illustrated. The parameters are $\Gamma=0.025\,$eV, $T=0.1\,$eV, $\Phi=0\,$eV, $\varepsilon_0=0\,$eV, and $A_\text{D}=0.2\,$eV. }
    \label{fig:example_limit_cycles_non-interacting}
\end{figure}

To allow for a fair comparison of the device operation in different parameter regimes, we characterize the limit cycle dynamics
of an observable $\av{O}(t)$ by its cycle average $\overline{\av{O}}$, amplitude $\Delta \av{O}$, and phase shift $\Delta \varphi_{\av{O}}$ in comparison to the adiabatic response limit, i.e., the limit of $\Omega_\text{D}\to 0\,$eV. 
We define the phase shift by the behavior of the fundamental frequency in the discrete Fourier decomposition of the observable dynamics,
\begin{align}
    \Delta \varphi_{\av{O}} = - i\, \text{ln}\left(\frac{a_1}{|a_1|} \right) - \Delta \varphi_{\av{O}}^{(\Omega_\text{D}\to 0)},
\end{align}
where we subtract the phase shift $\Delta \varphi_{\av{O}}^{(\Omega_\text{D}\to 0)}$ corresponding to the limit $\Omega_\t{D}\to 0$ and $a_k$ denote the frequency components obtained from a discrete Fourier decomposition of $N$ data points $(t_j,\av{O}_j)$
\begin{align}
    a_k = \frac{1}{N} \sum_{j=0}^{N-1} \expo{- i k \Omega_\text{D} t_j } \av{O}_j.
\end{align}
In Fig.~\ref{fig:example_limit_cycles_non-interacting}, these quantities are illustrated using the limit cycle dynamics of the electronic population $\elpop$ for different driving frequencies $\Omega_\text{D}$. The amplitude, $\Delta \elpop$, is obtained from the difference between the maximum and minimum during the limit-cycle dynamics.
Moreover, the phase shift of the fundamental frequency, $\Delta \varphi_{\elpop} = \nicefrac{\Delta t_{\elpop}}{\Omega_\text{D}}$, is visualized via a delay in time, $\Delta t_{\elpop}$, of the maximum of the according Fourier component.

The list of considered observables includes the electronic population $\elpop$, the charge current $I$, the displacement of the harmonic mode $\pos$, the occupation of the harmonic mode $\vibpop$, and the driving-field-induced power $\power$. 
In the case of the periodic driving field introduced in Eq.~\eqref{eq:Hamiltonian}, the induced power is given by
\begin{align}
    \power = A_\text{D}\Omega_\text{D} \text{cos}\left( \Omega_\text{D} t\right) \elpop(t).
\end{align}
Analogous to a mechanical resonator, the power absorption from the driving field is maximal for a $\frac \pi 2$-shifted response in the electronic population.

\section{Results and discussion}
\label{sec:interplay_driving_field_harmonic_mode}

In the following, we investigate the effect of the periodic driving field on an open quantum system exhibiting electronic-vibrational interaction. In Sec.~\ref{sec:Results:Equilibrium}, we present results without directed charge transport through the interacting system and without vibrational relaxation. Subsequently, we add layer by layer the influence of bias voltages (see Sec.~\ref{sec:Results:Bias}) and vibrational relaxation (see Sec.~\ref{sec:Results:VibRelax}).

\subsection{Application of the driving field to a system in equilibrium}
\label{sec:Results:Equilibrium}
We begin our investigation with the electronic-vibrational interacting system without an applied bias voltage, $\Phi=\mu_L-\mu_R=0$, and without vibrational relaxation, $\Lambda=0$. 

\begin{figure}[!b]
 \centering
 \footnotesize
    \includegraphics{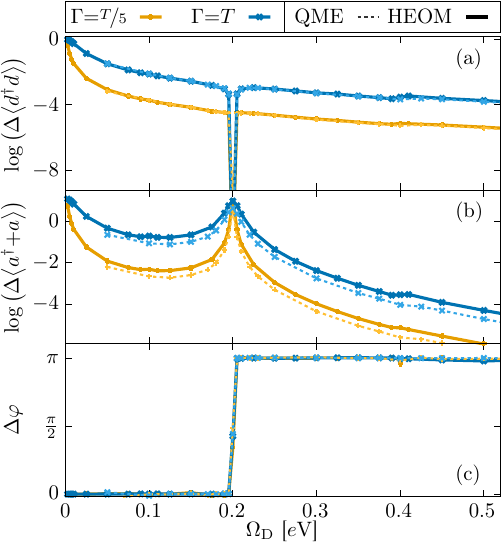}
    \caption{Amplitudes of (a) electronic population and (b) harmonic oscillator displacement dynamics as a function of the driving frequency. Furthermore, in panel (c) the relative delay of these dynamics is shown based on the corresponding phase difference $\Delta \varphi = \Delta \varphi_{\elpop} - \varphi_{\pos}$. Further parameters are $\overline{\varepsilon}_0=0\,$eV, $\Omega=0.2\,$eV, $\frac \lambda \Omega = 1.5$, $A_\text{D}=0.4\,$eV, and $T=0.025\,$eV.
    }
    \label{fig:amplitudes_elpop_n_displacement_wo_bias}
\end{figure}

Figures~\ref{fig:amplitudes_elpop_n_displacement_wo_bias}(a) and \ref{fig:amplitudes_elpop_n_displacement_wo_bias}(b) show the driving field induced dynamical amplitudes of both the electronic population and the displacement of the harmonic mode as a function of the driving frequency.
In the limit of slow driving frequencies, $\Omega_\text{D}\to 0$, the relaxation of the open quantum system is fast in comparison to the driving frequency. Hence, the state of the system is quasi-instantaneously adapting to the energetic situation, and the amplitude of the observables is determined by the amplitude of the observables in the instantaneous stationary state within the swept energy range. In the opposite limit, $\Omega_\text{D}\to\infty$, the driving frequency is faster than any relaxation process of the open quantum system. Accordingly, the system approaches the state corresponding to the average energy of the driving field and the amplitudes of all observables vanish. The general form of the transition between these limits is influenced by the different time scales of the relaxation processes like the coupling strength, $\Gamma$, and the temperature, $T$. 

At the resonance of the driving frequency with the frequency of the harmonic mode, $\Omega_\text{D}=\Omega$, we find additionally a collapse of the electronic population dynamics as well as a strong enhancement in the harmonic oscillator (HO) displacement dynamics.
We can understand this behavior in the context of the mechanical resonator. The harmonic oscillator absorbs energy very efficiently at its resonance frequency leading to pronounced oscillation in its displacement. The connection to the mechanical resonator is further supported by the phase shift between the dynamics of the electronic population, $\Delta \varphi$, which exhibits a $\pi$-step behavior at resonance (see Fig.~\ref{fig:amplitudes_elpop_n_displacement_wo_bias}(c)). 
However, the HO displacement also has a back-action on the electronic population, which serves as a driving force of the HO. Hence, the amplitude of the HO displacement dynamics can only increase until its influence on the electronic population dynamics is exactly counteracting the periodic driving field. Consequently, the electronic system is effectively decoupled from the driving field at resonance.

\begin{figure}[!t]
 \centering
 \footnotesize
 \includegraphics{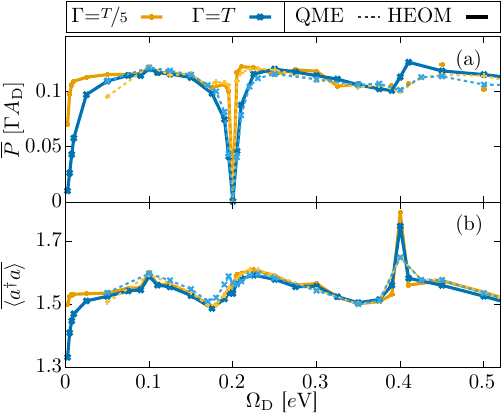}
 \caption{Limit-cycle-averaged vibrational excitation (a) and induced power (b) as a function of the driving frequency $\Omega_\text{D}$ for two different electronic system-environment couplings $\Gamma$. The dashed lines are QME results.  Further parameters are the same as in Fig.~\ref{fig:amplitudes_elpop_n_displacement_wo_bias}.} 
\label{fig:observable_average_vs_frequency_Amplitude0_4_Equilibrium_Eps0_00_el_ph1_5_different_Gamma}
\end{figure}

Another perspective of this decoupling mechanism is shown in Fig.~\ref{fig:observable_average_vs_frequency_Amplitude0_4_Equilibrium_Eps0_00_el_ph1_5_different_Gamma}(a), where the average driving power over a cycle induced by the driving field vanishes at resonance. 
Furthermore, in Fig.~\ref{fig:observable_average_vs_frequency_Amplitude0_4_Equilibrium_Eps0_00_el_ph1_5_different_Gamma}(b) we plot the cycle averaged vibrational excitation of the harmonic oscillator as a function of the driving frequency. 
In agreement with the effective decoupling from the driving field, the vibrational excitation does not exhibit a sharp resonance feature.
Instead, we observe a relatively broad and moderate increase of the vibrational excitation in the vicinity of the resonance frequency. 
It reflects the quantized energy transfer assisted by the driving field that provides the necessary energy for charge transfer processes involving one vibrational quantum of the harmonic oscillator~\cite{Tien1963,Kuperman2020}. Similar to the step-wise increase of the charge current with increasing bias voltage, the vibrational excitation is correspondingly enhanced when the driving frequency allows for $n$-phonon processes at the frequencies $\Omega_\text{D} = n\Omega$.
Moreover, the small peak at $\Omega_\text{D} = \frac \Omega 2$ is another fingerprint of the quantized interaction with the driving field as it results from a conversion of two driving field quanta to one excitation of the harmonic oscillator.
Furthermore, the driving field causes sharp peaks in the vibrational excitation at $\Omega_\text{D} = 2\Omega$. 
Such a peak is known from a classical parametric resonance, which can also be exploited on a swing. 
In contrast to the resonance case, the decoupling mechanism mediated by the electronic-vibrational interaction is not observed at the driving frequencies of both sharp peaks, $\Omega_\text{D} = \frac \Omega 2$ and $\Omega_\text{D} = 2\Omega$.

The results of QME and HEOM are consistent in both strengths of $\Gamma$. Although $\Gamma=T$ is beyond the weak coupling region, the electronic-vibrational interaction reduces the effective coupling strength\cite{Eidelstein2013} giving rise to the reliability of QME.

\subsection{Effects of an applied bias voltage}
\label{sec:Results:Bias}
So far, we have focused on the action of the driving field on a system without directed transport through the nanosystem. Next, we consider the additional effect of an applied bias voltage $\Phi=\mu_L-\mu_R$. For small bias voltages, charge transport is suppressed due to the electronic-vibrational interaction, an effect also referred to as the Franck-Condon blockade.\cite{Mitra2004,Koch2005}
In the following, we add the charge current to the observables of interest and focus on the induced power as a reference for the effective decoupling mechanism.

\begin{figure}[!b]
 \centering\includegraphics{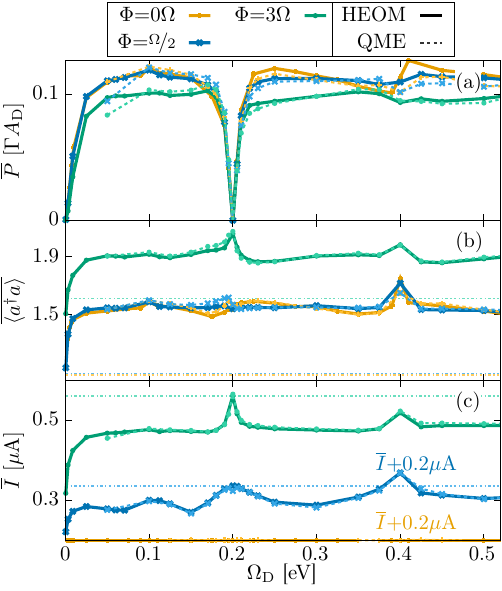}
     \vspace{-0.2cm}
\caption{Shown are the limit-cycle-averaged induced power (a), vibrational excitation (b), and charge current (c) as a function of the driving frequency $\Omega_\text{D}$ for two non-vanishing bias voltages $\Phi$  in comparison to the zero bias voltage case. For an improved comparison of the driving-frequency dependence, some data sets are shifted by a constant value specified next to the data. The light dashed lines indicate the result in the limit of a vanishing driving amplitude. The lighter colored points are QME results. 
 Further parameters are $\overline{\varepsilon}_0=0\,$eV, $A_\text{D}=0.4\,$eV, $\frac{\lambda}{\Omega}=1.5$, $\Omega=0.2\,$eV, and $T\,{=}\,\Gamma=0.025\,$eV. }
    \vspace{-0.2cm} \label{fig:observable_average_vs_frequency_Amplitude0_4_different_Bias_Eps0_00_el_ph1_5}
\end{figure}

Figure~\ref{fig:observable_average_vs_frequency_Amplitude0_4_different_Bias_Eps0_00_el_ph1_5} illustrates the response of the open quantum system for three different bias voltages based on the induced driving power, the vibrational excitation, and the charge current through the junction as a function of the driving frequency.
The induced power exhibits a big drop at the resonance frequency of the harmonic oscillator for all bias voltages. 
Hence, the charge transport through the nanosystem does not interfere with the decoupling mechanism arising from the interplay of the electronic population and harmonic oscillator dynamics.
Moreover, we observe a reduction in the induced power once the bias voltage energetically allows for inelastic charge transport processes accompanied by excitations of the harmonic oscillator.
At these bias voltages, processes exciting the harmonic oscillator become energetically accessible whereas processes lowering the excitation of the harmonic oscillator become prohibited.\cite{Haertle2011II,Haertle2018}
Hence, the applied bias voltage significantly contributes to the excitation of the vibrational mode.
Consequently, the driving field competes against the bias voltage to induce power into the vibrational mode.
This is also reflected in the vibrational excitation (see Fig.~\ref{fig:observable_average_vs_frequency_Amplitude0_4_different_Bias_Eps0_00_el_ph1_5}(b)). 
Although the vibrational excitation 
generally increases as the bias voltage is increased, the largest relative enhancement is not obtained for the largest bias voltage values. Meaning, the enhancement with respect to zero driving field, i.e. $A_\text{D}\to 0$, is more significant for smaller values of the bias-voltage.

The decoupling mechanism is also expressed in the charge current which attains the same value as in the limit of no driving field when in resonance.
The peaked recovery of the charge current in combination with a remainder of the general driving-induced vibrational excitation also leads to a peak in the vibrational excitation for high bias-voltages.
 Moreover, the pronounced vibrational-excitation enhancement at $\Omega_\text{D}=2\Omega$ yields a lift of the Franck-Condon blockade~\cite{May2008II}. Accordingly, we find a charge-current enhancement beyond the limit without driving the field  up to intermediate bias voltages. 

Regardless of the bias, QME and HEOM give accordant results, which demonstrates the small broadening effect within these conditions. Moreover, the resonance features are not apparent wihtout electronic-vibrational interaction in the system (see Appendix \ref{sec:Appendix}).

\subsection{Influence of vibrational relaxation}
\label{sec:Results:VibRelax}

\begin{figure}[!t]
\centering
\includegraphics{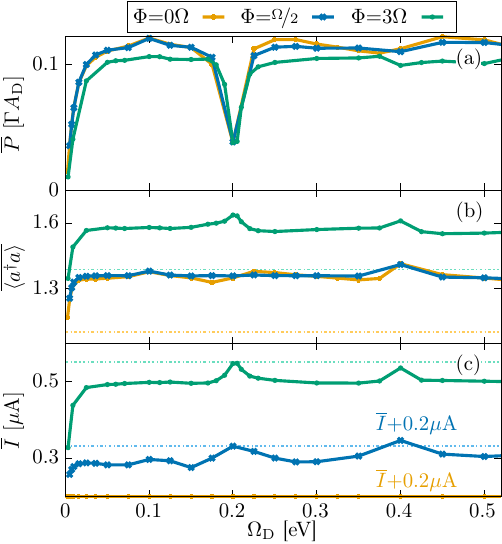}
    \vspace{-0.3cm}
 \caption{Shown are the limit-cycle-averaged induced power (a), vibrational excitation (b), and charge current (c) as a function of the driving frequency $\Omega_\text{D}$ for two non-vanishing bias voltages $\Phi$  in comparison to the zero bias voltage case. For an improved comparison of the driving-frequency dependence, some data sets are shifted by a constant value specified next to the data.  The light dashed lines indicate the result in the limit of a vanishing driving amplitude. Further parameters are $\overline{\varepsilon}_0\,{=}\,0\,$eV, $A_\text{D}\,{=}\,0.4\,$eV, $\frac{\lambda}{\Omega}\,{=}\,1.5$, $\Omega\,{=}\,0.2\,$eV,  $T\,{=}\,\Gamma=0.025\,$eV and $\Lambda\,{=}\,0.01\,$eV. }
    \vspace{-0.3cm} \label{fig:observable_average_vs_frequency_Amplitude0_4_different_Bias_Eps0_00_el_ph1_5_Lambda0_01}
\end{figure}

Finally, we explore the influence of vibrational relaxation on the decoupling mechanism as well as the lifting of the Franck-Condon blockade.
To this end, we show  the response of the open quantum system for three different bias voltages and comparatively strong environmental damping based on the induced power, the vibrational excitation, and the charge current through the junction as a function of the driving frequency in Fig.~\ref{fig:observable_average_vs_frequency_Amplitude0_4_different_Bias_Eps0_00_el_ph1_5_Lambda0_01}.

Including the vibrational relaxation effects, we notice that the collapse of the induced power at the resonant frequency is no longer complete (see Fig.~\ref{fig:observable_average_vs_frequency_Amplitude0_4_different_Bias_Eps0_00_el_ph1_5_Lambda0_01}(a)). 
In this case, the damping of the harmonic oscillator dynamics interferes with the effective decoupling mechanism. 
The relative enhancement of the vibrational excitation is weakened by the environment. This can be observed by comparing Fig.~\ref{fig:observable_average_vs_frequency_Amplitude0_4_different_Bias_Eps0_00_el_ph1_5_Lambda0_01}(b) and  Fig.~\ref{fig:observable_average_vs_frequency_Amplitude0_4_different_Bias_Eps0_00_el_ph1_5}(b).
The charge current is also affected by the environment and its profile round out.
However, the partial removal of the Franck-Condon blockade due to the vibrational excitation enhancement still remains visible in the charge current at $\Omega_\text{D}=2\Omega$
(see Fig.~\ref{fig:observable_average_vs_frequency_Amplitude0_4_different_Bias_Eps0_00_el_ph1_5_Lambda0_01}(c)). Moreover, the charge current still approximately recovers the limit of vanishing driving amplitude at  $\Omega_\text{D}=2\Omega$. These features hold true even for a comparatively strong damping strength, that is $\Lambda = \frac{5\Gamma}{2}$.

\section{Conclusions}
\label{sec:Conclusion}

In this paper, we investigated an electronic-vibrationally interacting system under the influence of a time-periodic driving field.
Employing the numerically exact HEOM approach, we examined the entire range from slow to fast driving frequencies.
Over a wide range of driving frequencies we observed a transition from an adiabatic slow driving case to a fast and quasistatic average energy case.
Beyond the limit-cycle-averaged observables, the amplitude and phase shift definitions facilitate a comprehensive picture of the driving induced dynamics. 
Overall, the study provides basic insights for engineering heat engines or machines based on quantum systems.

The periodically driven and electronic-vibrationally interacting system exhibits a nontrivial interaction between the harmonic oscillator and the driving field.
At resonance, $\Omega_\text{D}=\Omega$, we found that the influence of the harmonic oscillator on the electronic system is counteracting the driving field. 
Thus, a driving-frequency specific collapse of the response occurs in the electronic population and induced power. 
The application of a bias voltage is not affecting this mechanism. Accordingly, the charge-current response at resonance is independent of the driving amplitude.
At higher harmonics, the vibrational excitation and the charge current exhibit clear fingerprints of the interaction between the harmonic oscillator and the driving field.
In particular, at $\Omega_\text{D}=2\Omega$, a significant enhancement of the vibrational excitation causes a lifting of the Franck-Condon blockade. The introduction of an additional environmental damping affects the interaction between the harmonic oscillator and the driving field. In particular, the harmonic oscillator counteraction to the periodic driving is no longer perfect.
Moreover, we found that the charge-current enhancement due to the driving induced lifting of the Franck-Condon blockade is persistent under the influence of environmental damping.

\begin{acknowledgments}
It is a pleasure to acknowledge fruitful discussions with C. Kaspar and S. Rudge.
WD acknowledges support by National Natural Science Foundation of China (NSFC No. 22273075). 
This work was supported by the German Research Foundation (DFG) through a research grant and FOR 5099.
Furthermore, support by the state of Baden-W\"urttemberg through bwHPC and the DFG through Grant No. INST 40/575-1 FUGG (JUSTUS 2 cluster) is gratefully acknowledged.
This research was supported by the ISRAEL SCIENCE FOUNDATION (Grant No. 1364/21).
\end{acknowledgments}

\cleardoublepage
\appendix 
\section{Driven open quantum system without vibrational degrees of freedom}
\label{sec:Appendix}

\begin{figure}[!b]
 \centering\includegraphics{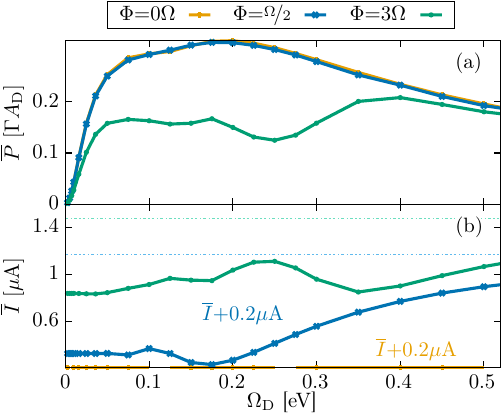}
     \vspace{-0.2cm}
\caption{Response of the open quantum system without electronic-vibrational interaction. Shown are the limit-cycle-averaged induced power (a), and charge current (b) as a function of the driving frequency $\Omega_\text{D}$ for two non-vanishing bias voltages $\Phi$  in comparison to the zero bias voltage case. For an improved comparison of the driving-frequency dependence, some data sets are shifted by a constant value specified next to the data. The light dashed lines indicate the result in the limit of a vanishing driving amplitude. Further parameters are $\varepsilon_0=0\,$eV, $A_\text{D}=0.4\,$eV, $\lambda=0$, and $T\,{=}\,\Gamma=0.025\,$eV. }
    \vspace{-0.2cm} \label{fig:observable_average_vs_frequency_Amplitude0_4_different_Bias_Eps0_00_el_ph0}
\end{figure}

We show the response of a time-periodically and externally driven open quantum system without vibrational degrees of freedom in Fig.~\ref{fig:observable_average_vs_frequency_Amplitude0_4_different_Bias_Eps0_00_el_ph0}.
The driving-frequency dependence of the induced power and the charge current are both reflecting the importance of the Floquet replica of the electronic state.~\cite{Tien1963,Moskalets2002,Kohler2005,Kuperman2020} 
In particular, the charge current reflects the de-/activation of the different transport channels in accordance with the Bessel functions of the first kind, $J_n\left(\frac{A_\text{D}}{\Omega_\text{D}}\right)$. 
However, the visible features are distinct from the resonances observed in the electronic-vibrationally interacting system. Hence, the features in the main text are related to the electronic-vibrational interaction.

\bibliography{./lit}

\end{document}